# Dual-tuned Coaxial-transmission-line RF coils for Hyperpolarized $^{13}$C and Deuterium $^{2}$H Metabolic MRS Imaging at Ultrahigh Fields

Komlan Payne, *Graduate Student Member, IEEE,* Yunkun Zhao, *Student Member, IEEE*, Aditya Ashok Bhosale, *Student Member, IEEE*, and Xiaoliang Zhang, *Member, IEEE*

*Abstract*— **Objective:** Information on the metabolism of tissues in healthy and diseased states plays a significant role in the detection and understanding of tumors, neurodegenerative diseases, diabetes, and other metabolic disorders. Hyperpolarized carbon-13 magnetic resonance imaging ($^{13}$C-HPMRI) and deuterium metabolic imaging ($^{2}$H-DMI) are two emerging X-nuclei used as practical imaging tools to investigate tissue metabolism. However due to their low gyromagnetic ratios ($\gamma_{13C}$ = 10.7 MHz/T; $\gamma_{2H}$ = 6.5 MHz/T) and natural abundance, such method required a sophisticated dual-tuned radiofrequency (RF) coil. **Methods:** Here, we report a dual-tuned coaxial transmission line (CTL) RF coil agile for metabolite information operating at 7T with independent tuning capability. The design analysis has demonstrated how both resonant frequencies can be individually controlled by simply varying the constituent of the design parameters. **Results:** Numerical results have demonstrated a broadband tuning range capability, covering most of the X-nucleus signal, especially the $^{13}$C and $^{2}$H spectra at 7T. Furthermore, in order to validate the feasibility of the proposed design, both dual-tuned $^{1}$H/$^{13}$C and $^{1}$H/$^{2}$H CTLs RF coils are fabricated using a semi-flexible RG-405 .086" coaxial cable and bench test results (scattering parameters and magnetic field efficiency/distribution) are successfully obtained. **Conclusion:** The proposed dual-tuned RF coils reveal highly effective magnetic field obtained from both proton and heteronuclear signal which is crucial for accurate and detailed imaging. **Significance:** The successful development of this new dual-tuned RF coil technique would provide a tangible and efficient tool for ultrahigh field metabolic MR imaging.

*Index Terms*—Brain glucose metabolisms, deuterium magnetic resonance spectroscopy, High/low-impedance RF coil, hyperpolarized $^{13}$C, metabolic imaging, ultra-high field.

This work was supported in part by the NIH under Grant U01 EB023829 and SUNY Empire Innovation Professorship. (Corresponding author: Xiaoliang Zhang.)

Komlan Payne is with Department of Biomedical Engineering, State University of New York at Buffalo, Buffalo, NY 14260 USA (e-mail: komlanpa@buffalo.edu).
Yunkun Zhao is with Department of Biomedical Engineering, State University of New York at Buffalo, Buffalo, NY 14260 USA (e-mail: yunkunzh@buffalo.edu).
Aditya Ashok Bhosale is with Department of Biomedical Engineering, State University of New York at Buffalo, Buffalo, NY 14260 USA (e-mail: adityaas@buffalo.edu).
Xiaoliang Zhang is with the Departments of Biomedical Engineering and Electrical Engineering, State University of New York at Buffalo, Buffalo, NY 14260 USA (e-mail: xzhang89@buffalo.edu).

## I. INTRODUCTION

METABOLIC pathway-based magnetic resonance spectroscopic imaging (MRSI) is a promising tool that can be used to reveal the proper functioning of a biological system [1-7]. This technique provides key information for the diagnosis and treatment monitoring of various diseases, including cancer, diabetes, and neurodegenerative disorders [8-13]. In contrast to MRSI, fluorodeoxyglucose positron emission tomography (FDG-PET), a common metabolic imaging tool, has found success in tumor detection and treatment monitoring in clinical studies by providing high-resolution maps of glucose uptake [14-17]. However, the use of radioactive contrast and the lack of direct information on glucose metabolism have limited its full potential [18, 19]. Conventionally, metabolic MRSI suffers from limited resolution due to its low sensitivity. Recent development demonstrated feasibility of using hyperpolarized carbon-13 magnetic resonance imaging ($^{13}$C- HPMRI) and deuterium metabolic imaging ($^{2}$H-DMI) to probe the tissue metabolic process in vivo, showing much improved sensitivity [20-26]. Unlike FDG-PET, these methods are non-radiative. While DMI method can be performed after oral administration of [6,6',-$^{2}$H$_2$]glucose, $^{13}$C- HPMRI method involved intravenous injection of hyperpolarized [1-$^{13}$C]pyruvate [21, 27]. Both imaging tools, although using different methodology, have proven to be robust and easy to perform in animal brain [28-30] which is further expended to human brain and liver [22, 24, 31-33]. Spectroscopic data acquisition of $^{13}$C- HPMRI and $^{2}$H-DMI at 3 Tesla has revealed insufficient inherent spectral resolution [27], leading to a growing interest in high and ultrahigh field studies [28, 34-39] for the proven SNR gain despite technical challenges [13, 40-53]. Due to the low natural abundance and nuclear spin polarization of these X-nuclei (non-hyperpolarized), a weak signal is expected, limiting their signal-to-noise ratio (SNR) and thus quantitative assessment of this metabolic imaging technique [27, 28, 34, 35, 54-56]. Previous studies have indicated that the SNR of the X-nucleus can be significantly enhanced using a dual-tuned RF coil at ultrahigh fields [34, 35, 57-62]. A single coil [63, 64] or two separate coils [57, 58, 65, 66] can both be used to implement a dual-tuned RF coil, owing to their strength and weakness. By adopting the two-coil design, the electromagnetic coupling between the coils can lead to the degradation of the $B_1$ field intensity. To address this issue, a decoupling network is often required to reduce the coupling effects between the two coils. However, the inclusion of a decoupling network introduces additional losses in the design, potentially affecting the overall efficiency of the coils. Alternatively, using an orthogonal-field coil design with a specific geometric configuration can help mitigate the coupling and improve the B1 field intensity. However, this approach can increase the complexity of the coil design [57, 58, 65, 67-69]. In a single-coil dual-tuned design, the coil is intended to be used for multiple resonant frequencies. One of the main challenges with this design is achieving independent tuning between the resonant frequencies. The "two-pole" technique using an LC trap circuit is implemented to design a dual-tuned single-port surface coil with similar $B_1$ field distribution for both the proton and the hetero nucleus [64]. The same technique is also used for double-tuned birdcage volume coil excited using a single port with double-tuned direct drive or inductive drive [70]. Another common challenge in dual-tuned coil designs is that the two fields of the proton and hetero-nucleus are often not completely overlapped. This makes $B_0$ shimming and structural-spectroscopic image correlation less efficient. In this work, we proposed a single-coil dual-tuned coaxial transmission line (CTL) RF coil with independent tuning capabilities and highly overlapped $B_1$ profile of the two nuclei for X-nuclear MRS imaging aimed at probing metabolite information.

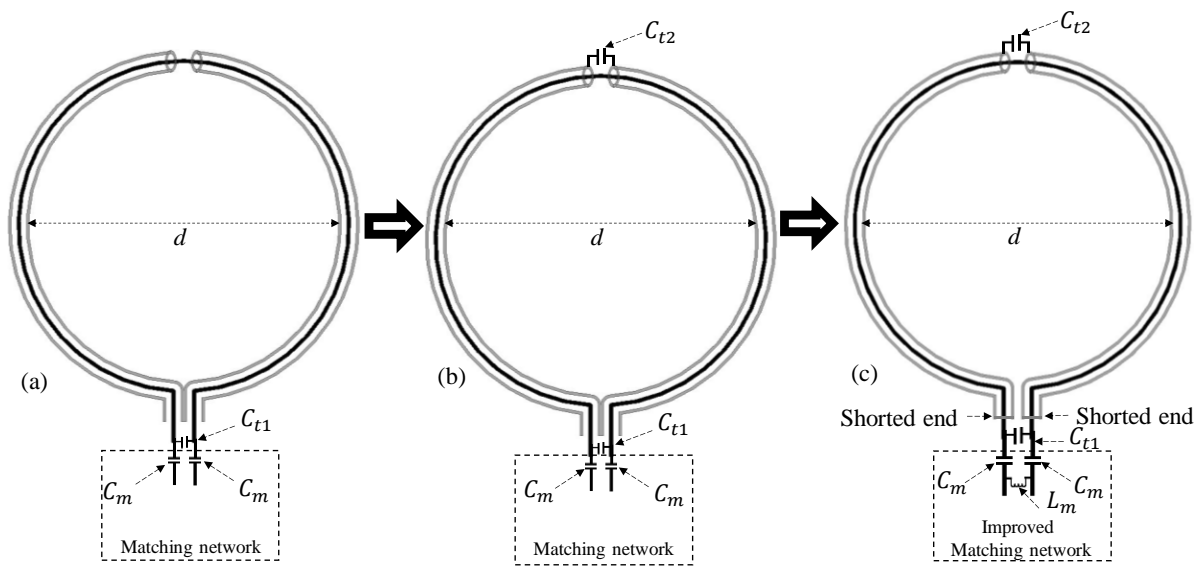

Fig. 1. Evolution process of the proposed dual-tuned shorted end CTL coils with independent tuning capabilities. (a) Conventional CTL RF coil for single tuned applications; (b) CTL coil with tuning capacitance ($C_{t2}$) at the gap of the outer shield for single tuned applications; (c) Proposed dual-tuned shorted end CTL with independent tuning capability

Recently, intensive research has been conducted on CTL RF coils for nuclear magnetic resonance (NMR) at frequencies ranging from high field to ultrahigh field [71-77]. The flexibility of CTL RF coils is to some extent needed to image dynamic anatomy of the human body, as it has lessened the burden of anatomical posture constraints. These types of RF coils are made of coaxial cable with the port connected to the inner conductor and loaded with a gap on the outer shield located opposite to the port [71]. Although they are not truly flexible since lumped components are also integrated within the design, such technology is inexpensive and can be adapted to variations in anatomy. The study of CTL RF coils has revealed multimode operating frequencies associated with the design parameters. While the fundamental frequency is often used to implement receiver RF array coils [71, 73], the second operating mode has a high impedance characteristic and can be used to design a tight-fitting transceiver array [72]. The multimode CTL can be used to implement a dual-tuned RF coil to study non-proton nuclei (X-nuclei) for magnetic resonance spectroscopy and imaging. By adding gaps in the inner layer or the outer layer, the resonant frequencies of the CTL can be tuned [78]. However, the resonant frequencies of the multimode CTL cannot be tuned independently to the desired frequencies as they are correlated with each other. In order to alleviate this issue, two asymmetric gaps is added on the outer conductor of the CTL [79] to achieve independent tuning of both bands. The location of the two asymmetric gaps in the outer shield can be carefully selected for the CTL coil through iterative optimization using a full-wave electromagnetic solver to operate at the targeted dual-band frequencies. While this technique allows independent tuning of the two resonant frequencies, in practice, the position of the asymmetric gaps in the outer shield cannot be easily changed to compensate for load variations or cable bending to retune the resonant frequencies.

In this work, we propose an alternative dual-tuned CTL RF

coil in which the first and second mode resonance frequencies can be tuned independently by simply varying the constituent of the design parameters using lumped components. In the next section, we elaborate on the development process of our proposed dual-tuned CTL with independent tuning capabilities, namely "shorted end coaxial transmission line" (SECTL). As a proof of concept, we demonstrate that both resonant frequencies can be controlled individually by simply varying the value of the lumped components in the design using full-wave simulation and bench test results. Furthermore, as an example, the proposed SECTL design is used to implement dual-tuned $^1$H/$^{13}$C and $^1$H/$^2$H RF coils for the ultrahigh field MR applications at 7T, which are valuable for structural and metabolic information detection. Finally, we investigate the efficiency and distribution of the magnetic field strength $B_{1,eff}$ of the proposed dual-tuned SECTL which indicate its feasibility for metabolic clinical applications.

## II. Description of Methodology

The evolution process of the proposed dual-tuned SECTL with independent tuning capabilities is depicted in Fig. 1. The process starts with the conventional CTL RF coil previously introduced [71-73], where $C_m$ is the matching capacitor and $C_{t1}$ is used for tuning the resonant frequencies [see Fig. 1(a)]. In such designs, the lumped tuning element $C_{t1}$ simultaneously controls both the first ($f_1$) and second ($f_2$) resonant frequencies, preventing independent tuning capability of the two operating bands. In Fig. 1(b), we introduced another tuning capacitor $C_{t2}$ added to the gap at the outer shield. Simulation analysis has demonstrated that both resonant frequencies ($f_1$ and $f_2$) are also controlled by the tuning capacitor $C_{t2}$. By isolating the two arms of the outer shielding from each other and shorting the inner conductor to the outer conductor [Fig. 1(c)], we obtained the proposed design namely "shorted end coaxial transmission line" (SECTL). Such alternative design provides independent tuning capability such that the resonant frequencies $f_1$ and $f_2$ are controlled by $C_{t1}$ and $C_{t2}$, respectively. It is worth mentioning that a pi impedance matching circuit is employed to achieve optimal matching at both frequencies. A proper dual-band matching network [70] could be used instead for accurate matching of both resonant frequencies but would require more lumped components which can add resistive losses in the overall design. The proposed SECTL design is used to implement a dual-tuned designed for 7T with independent tuning capabilities for X-nuclear MRS Imaging aiming for metabolite information. To validate the capability of the proposed dual-tuned SECTL with independent control of each resonant frequency, we evaluated the scattering matrix of the design for a fixed value of $C_{t1}$ while varying $C_{t2}$ and vice versa. The simulated surface current distribution on the proposed coil is also obtained at both resonant frequencies to analyze their current path on the conductors. Furthermore, the simulated/measured $B_{1,eff}$ fields mapping of the dual-tuned SECTL designed for carbon-13 magnetic resonance imaging ($^{13}$C- HPMRI) and deuterium metabolic imaging ($^2$H-DMI) are also obtained at their corresponding resonance frequencies.

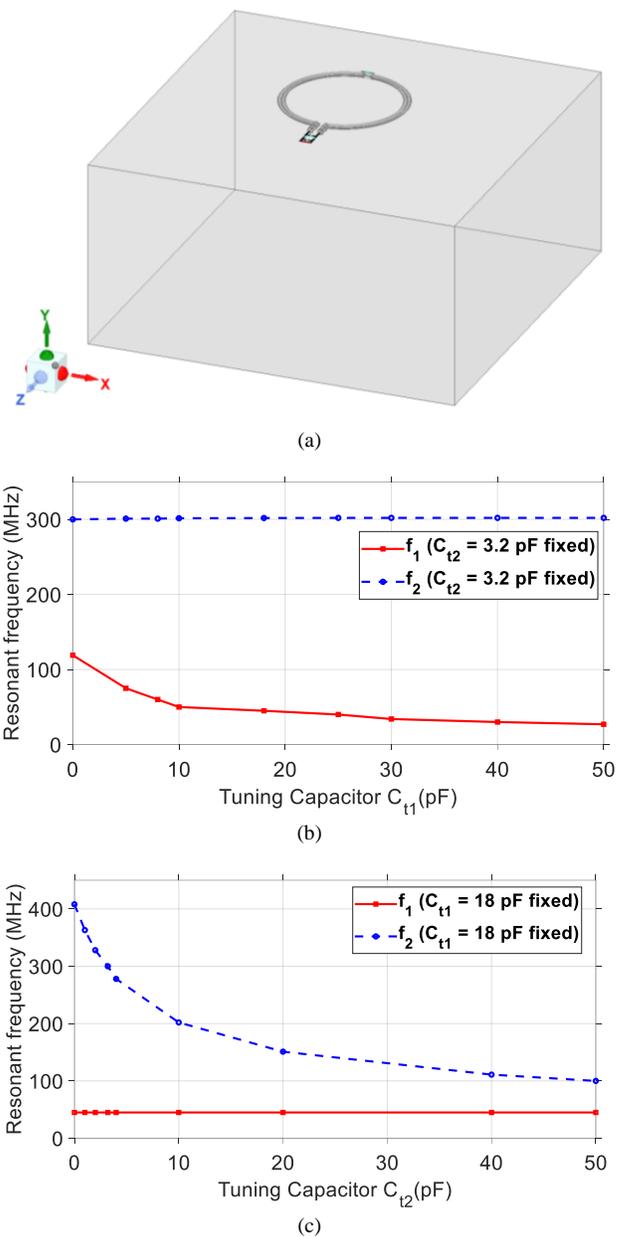

Fig. 2. (a) Proposed shorted end coaxial transmission line RF coil loaded with a cuboid phantom. Simulated frequency response result showing the effect of the parametric study. (b) The first resonant frequency $f_1$ is controlled by the tuning capacitor $C_{t1}$; (c) The second resonant frequency $f_2$ is controlled by the tuning capacitor $C_{t2}$. Results show that the two frequencies $f_1$ and $f_2$ can be tuned independently by $C_{t1}$ and $C_{t2}$, respectively.

## III. Proof of the Proposed Dual Tuned SECTL With Independent Tuning Capability

### A. Parametric study results

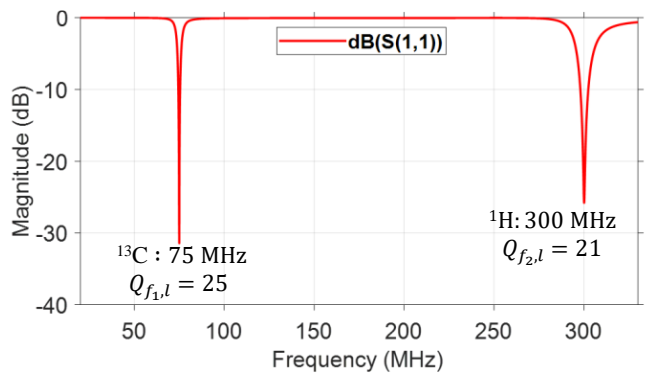

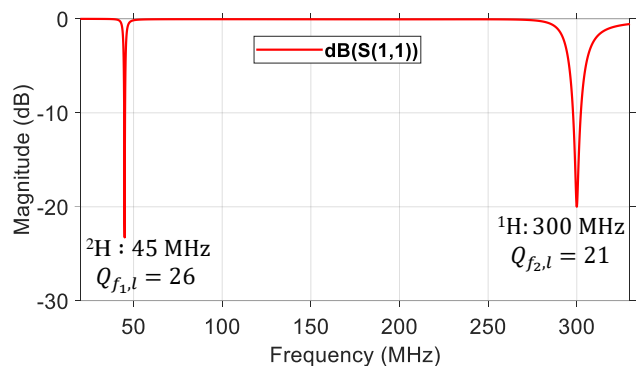

Fig. 3. Simulated S-parameter of the dual-tuned SECTL RF coil loaded with the cuboid phantom. (a) Results obtained for the $^1$H/$^{13}$C design where $C_{t1}$ = 6 pF, $C_{t2}$ = 3.2 pF, $C_m$ = 22 pF, and $L_m$ = 67 nH. (b) Results obtained from the $^1$H/$^2$H design where $C_{t1}$ = 18 pF, $C_{t2}$ = 3.2 pF, $C_m$ = 16 pF, and $L_m$ = 74 nH.

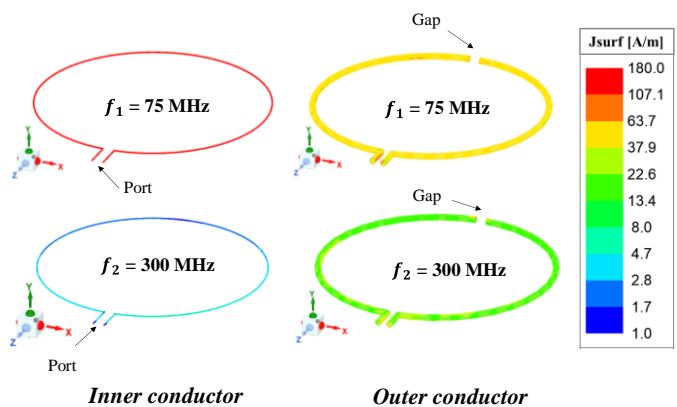

Fig. 4. Simulated surface current distribution in 3D plot of the proposed dual-tuned $^1$H/$^{13}$C RF coil operating at 7 Tesla. The primary current resides at the inner conductor for the low resonant frequency ($f_1$ = 75 MHz) while for the high resonant frequency ($f_2$ = 300 MHz), the main current appeared on the outer conductor.

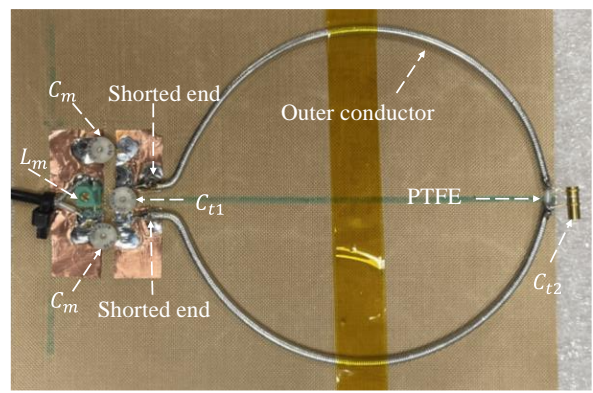

Fig. 5. Photograph of the constructed SECTL dual-tuned RF coils with tuning capacitor and inductor. The fabricated RF coil is independently tuned for $^1$H/$^{13}$C or $^1$H/$^2$H at 7 Tesla.

The proposed SECTL is placed 0.5 cm on top of a tank phantom with dimension 20 x 10 x 20 cm$^3$. The conductivity value $\sigma$ = 0.6 S/m and permittivity value $\varepsilon_r$ = 50 are used in the simulation model of the phantom to imitate the human brain tissue properties [see Fig. 2(a)]. The electrical parameters of the material used for the SECTL are the same as the commercial "RG-405 .086" coaxial cable. The diameter of the RF coil is roughly about 8-cm. A full wave simulation using High-Frequency Structure Simulator is used as design analysis for the dual-tuned RF coil. The coil is initially optimized for the second resonant frequency $f_2$ to operate at 300 MHz, the Larmor frequency of the proton $^1$H at 7T ($C_{t2}$ = 3.2 pF). Then, a parametric study is performed on the SECTL by simply varying the value of $C_{t1}$ while other parameters are fixed. The simulations results show that $f_1$ can be independently tuned in a wide range of frequencies (27 MHz – 122 MHz) while $f_2$ remains fixed around 300 MHz [see Fig. 2 (b)]. Note that $f_1$ covers most of the X-nuclear signal (such as $^{31}$P, $^{13}$C, $^{23}$Na, and $^2$H) at 7T. On the other hand, by fixing the value of $C_{t1}$ = 18 pF (such as $f_1$ operates at 45 MHz, the deuterium spectrum at 7T), while varying the value of $C_{t2}$, the second resonant frequency $f_2$ also shows independent tuning and cover a broad range of spectrum from 100 MHz to 408 MHz [see Fig. 2 (c)].

### B. Design of dual-tuned RF coils

Based on the results obtained from the parametric study, dual-tuned RF coils for metabolites quantification are feasible. Hence, dual-tuned $^1$H/$^{13}$C and $^1$H/$^2$H SECTLs RF coils are designed for hyperpolarized carbon-13 MRI and deuterium metabolic imaging (DMI), respectively. Both designs are tuned to their corresponding spectrum and matched to 50 ohms using co-simulation methods. The simulated scattering parameters of the dual-tuned RF coils are shown in Fig. 3. Results indicate a dual-tuned ($^1$H/$^{13}$C) response at 7T where the Larmor frequencies of the proton $^1$H and the X-nucleus $^{13}$C operate at 300 MHz and 75 MHz, respectively. As for the dual-tuned ($^1$H/$^2$H), the deuterium $^2$H operates at 45 MHz. A good matching is obtained at both frequencies for all the designs using the simple pi-matching network. The quality (Q) factor is analytically obtained from the ratio of the resonant frequency to the -3dB bandwidth. As shown in Fig.3, the simulated loaded quality factors at $^{13}$C, $^2$H and $^1$H are 25, 26 and 21, respectively. The Q-factor values of our proposed dual-tuned coaxial transmission line RF coil are pretty low compared to conventional single or dual-tuned single-conductor loop RF coils [63, 80]. Possibly, low Q is a drawback of coaxial transmission line RF coil [81].

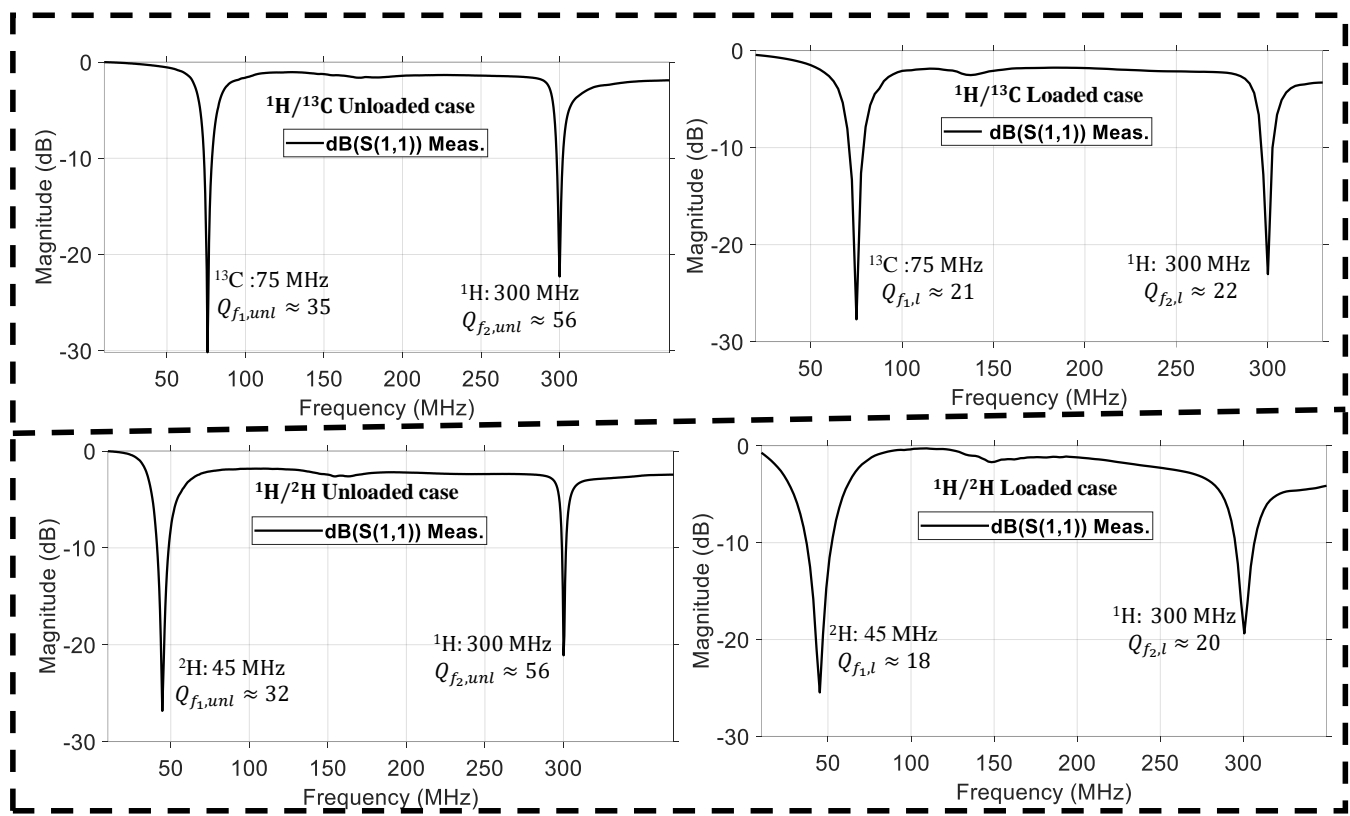

Fig. 6. Measured scattering parameters of the fabricated SECTL dual-tuned RF coil for both unloaded and loaded cases. The coil is appropriately tuned at 7T for dual-tuned $^1$H/$^{13}$C operation where the Larmor frequencies of the proton $^1$H and the X-nucleus $^{13}$C operate around 300 MHz and 75 MHz (a) and for dual-tuned $^1$H/$^2$H operation where the Larmor frequencies of the proton $^1$H and the X-nucleus $^2$H operate around 300 MHz and 45 MHz, respectively.

The surface current distribution on the conductors of the proposed dual-tuned ($^1$H/$^{13}$C) coaxial-transmission-line RF coil is obtained through full wave simulation at both operating bands ($f_1$ = 75 MHz and $f_2$ = 300 MHz). Simulated results illustrated in Fig. 4 show different current distributions on the coil conductors for both resonant frequencies. A uniform and strong surface current magnitude is observed at the inner conductor for the low resonant frequency $f_1$. As for the high resonant frequency $f_2$, the main current density resides in the outer shield of the coaxial transmission line coil with constant magnitude. Note that at the inner conductor, the current flow is not uniform at the high resonant frequency $f_2$, having high and low current density at the feeding and the gap of the shield, respectively.

*C. Fabrication and measurement results*

A dual-tuned SECTL RF coil adept to be tuned for $^1$H/$^{13}$C or $^1$H/$^2$H at 7 Tesla is constructed using the commercial semi-flexible "RG-405 .086" coaxial cable. The inner conductor (with 0.56 mm diameter) and the outer conductor (with 2.2 mm diameter) are isolated using a solid extruded PTFE insulation dielectric material (with 1.7 mm outer diameter, a relative permittivity value of $\varepsilon r$ =2.5 and a loss tangent of $\tan\delta$ = 0.001). At the two ends of the coaxial cable, a small section of the outer conductor and the insulation is removed. Then, the exposed inner conductor is bent and soldered together with the outer conductor. The physical dimensions of the RF coils are the same used in the numerical simulations. Variable capacitors and inductors integrated to the RF coils are soldered at their specific location for matching and tuning purposes. The fabricated dual-tuned RF coil is shown in Fig. 5. A variable trimmer capacitor Johanson 9615 ranging from 5 pF – 25 pF is used for $C_{t1}$ and $C_m$ while Johanson 27271 capacitor ranging from 0.6 - 4.5 pF is used for $C_{t2}$. Tunable RF inductor from Coilcraft with inductance value ranging from 65 – 99 nH is used for $L_m$. For the loaded case, the fabricated coil is placed on top of a 20 x 10 x 20 cm$^3$ tank containing 51.1% of sucrose in water doped with 0.25 gl$^{-1}$ NaCl. The phantom has roughly the same electrical parameter used in the numerical simulation. The scattering parameters of the fabricated RF coil are obtained using a vector network analyzer (from KEYSIGHT, E5061B 100 KHz -3 GHz, Santa Clara, CA, USA) properly calibrated to reduce ripples in the measurement data. For both unloaded and loaded case scenarios, the fabricated SECTL RF coil is successfully tuned and matched for $^1$H/$^{13}$C or $^1$H/$^2$H dual-tuned operation as shown in Fig. 6. The measured quality (Q) factor is estimated from the ratio of the resonant frequency to the -3dB bandwidth for both unloaded (Q-unloaded) and loaded (Q-loaded) case. For the $^1$H/$^{13}$C dual channel, the measured Q-ratio (unloaded/loaded) is 1.6 (35/21) for the $^{13}$C frequency and 2.5 (56/22) for the $^1$H frequency as illustrated in Fig. 6. Also, for the $^1$H/$^2$H dual channel, the measured Q-ratio is 1.7 (32/18) for the deuterium $^2$H frequency. The discrepancy (more pronounced at the lower frequencies) between the simulated and measured result is due to the fabrication tolerance, the loading effect, and the inherent losses within the lumped components as well as the loss from the feeding cable. From the measured result, it should be noted that lower Q-ratio is obtained at the X-nucleus ($^{13}$C or $^2$H) compared to the one observed at the proton ($^1$H). This situation reflects less coupling between the RF coil and the load at the lower frequency compared to the higher frequency due to the outer conductor. For future work, we believe that this issue can be solved by adding more gaps to the outer conductor of the RF coil which also will benefit the implementation of dual-tuned array for parallel imaging.

The simulated and measured magnetic field efficiency ($B_{1,eff}$) generated by the proposed SECTL RF coils are compared at the transversal and coronal planes in a vacuum (unloaded case). The field map is obtained for a 20 x 10 cm$^2$ central axial slice and a 10 x 10 cm$^2$ coronal slice located 0.5 cm away from the RF coil. The simulated $B_{1,eff}$ is obtained from the HFSS solver field calculator by normalizing the magnitude of $B_1$ field to 1 W of accepted power $(|B_1|/\sqrt{P_{acc}})$ considering the reflected power due to impedance mismatch at the port of the RF coils. As for the measured $B_{1,eff}$ field, a near field measurement technique similar to the one described in [82] is adopted. As shown in Fig. 7, a sniffer (H-field probe) is used to detect the B$_1$ field generated by the dual-tuned RF coil in 3D space. Such sniffer is integrated to a high-resolution router machine (Genmitsu CNC PROVerXL 4030) for accurate control and positioning of the field probe connected to a vector network analyzer (VNA) from Keysight, E5061B, Santa Clara, CA, USA. The resolution of the router machine is set 0.5 mm to ensure the accuracy of the field map. A matrix of 400 x 200 size is obtained for the transversal slice while 200 x 200 matrix size is obtained for the coronal slice. The raw data obtained from the VNA is processed using MATLAB to compute the measured $B_{1,eff}$ mapping in different orientations (both in transversal and coronal plane for this experiment) considering cable losses and input impedance matching. The data processing flow used to obtain the $B_{1,eff}$ mapping is shown in Fig. 8. Detailed information on the measurement $B_{1,eff}$ mapping obtained from the VNA raw data is included in the supporting document.

As can be seen in Fig. 9, excellent qualitative and quantitative agreement is obtained between the simulated and the measured $B_1^+$ mapping in both orientations except for some slight deviation due to the fabrication tolerance and the resolution of the router machine. The dual-tuned $^1$H/$^{13}$C RF coil delivers similar field homogeneity [see Fig. 7(a)] at both frequencies ($f_1$ = 75 MHz, $f_2$ = 300 MHz) which is beneficial for data acquisition of the metabolite information while using the proton for structural imaging and shimming. Likewise, as illustrated in Fig. 7(b), comparable field distribution is also obtained for the dual-tuned $^1$H/$^2$H RF coil at its corresponding frequencies $f_1$ = 45 MHz and $f_2$ = 300 MHz. For both RF coils, it can be pointed that, at the lower operating frequency ($f_1$), stronger magnetic flux density is obtained compared to the one obtained at higher frequency ($f_2$). This is due to the increase of radiation losses at higher operating frequency from the coaxial transmission line. Nevertheless, the simulated $B_{1,eff}$ field efficiency shown that the dedicated dual-tuned RF coils are feasible and potential candidate for high resolution metabolite images.

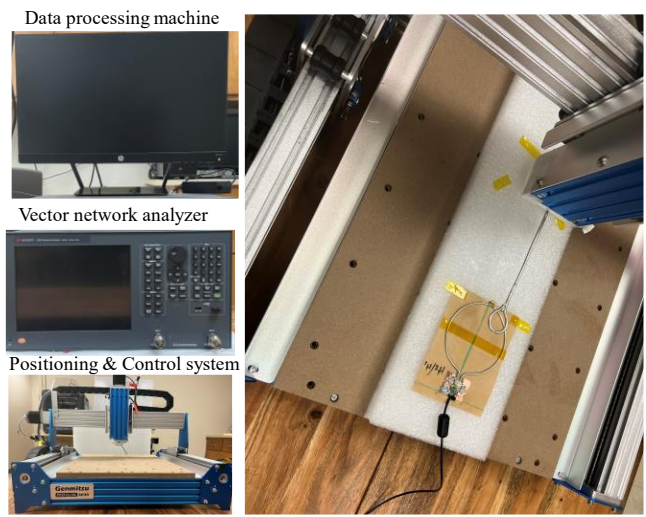

Fig. 7. Photograph of the near field measurement setup including the H-field probe, the milling machine, the vector network analyzer, and the data processing machine.

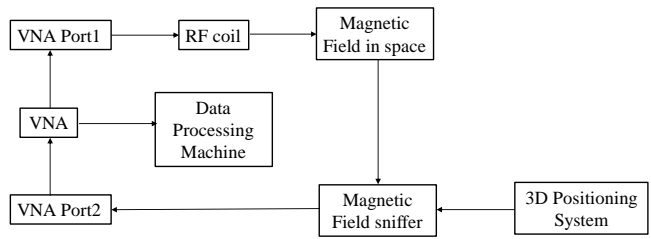

Fig. 8. Illustration of the data processing flow used to obtain the $B_{1,eff}$ mapping. The specification of the data processing machine is as follow: Processor: 11th Gen Intel(R) Core(TM) i7-11700 @ 2.50GHz; Installed RAM: 128.0 GB; GPU: Nvidia GeForce RTX 3070.

IV. DISCUSSION

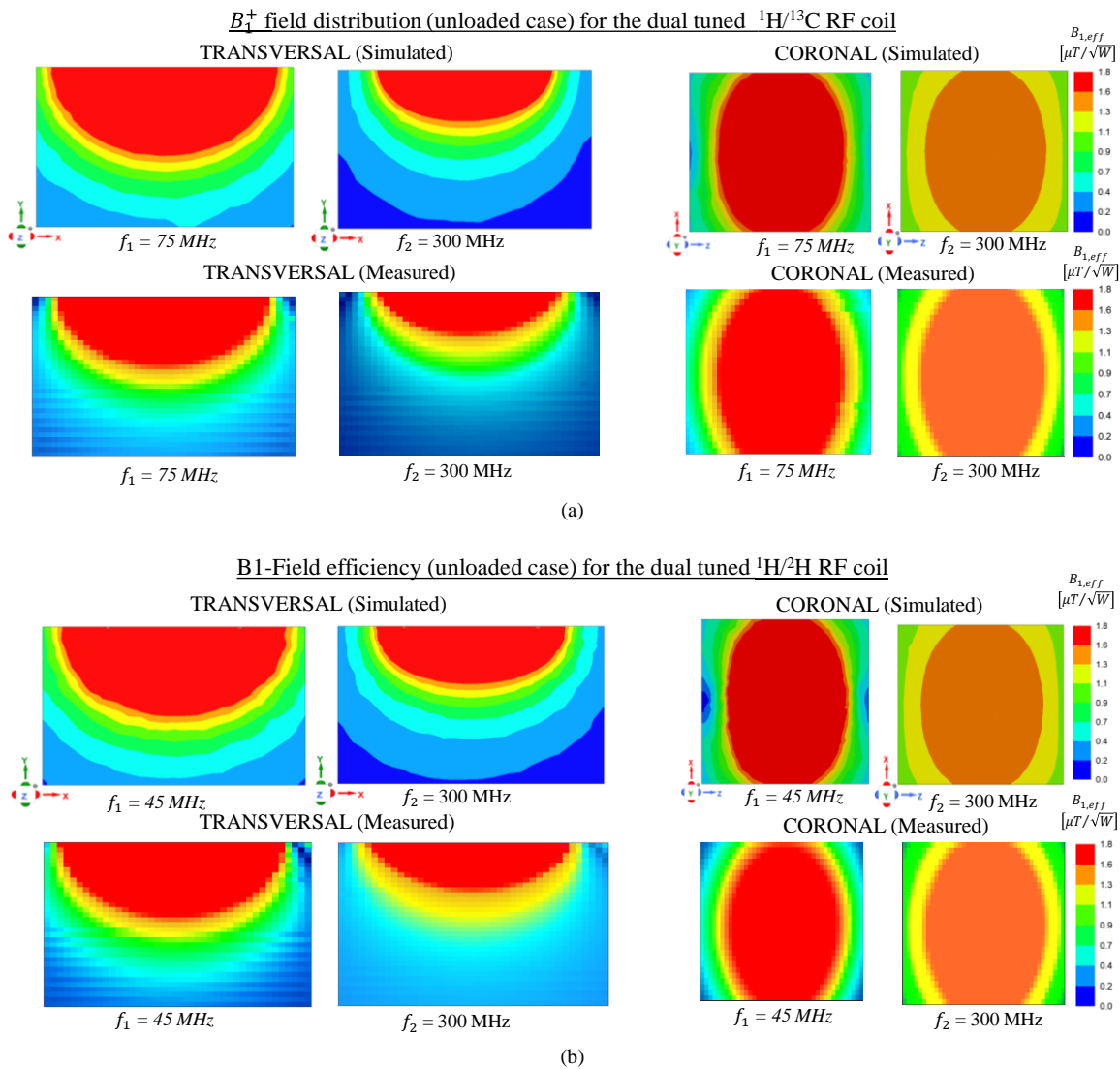

Fig. 9. Simulated and measured $B_{1,eff}$ field distribution (unloaded case) normalized to the accepted input power obtained for both axial and coronal plane; The experimental $B_1^+$ map is obtained for the 20 x 10 cm$^2$ central axial slice and a 10 x 10 cm$^2$ coronal slice 0.5 cm away from the RF coil. (a) Field obtained for the dual-tuned $^1$H/$^{13}$C RF coil at 7T; (b) Field obtained for the dual-tuned $^1$H/$^2$H RF coil at 7T. Good agreement is obtained between the simulated and measured field maps.

The proposed dual-tuned RF coaxial-transmission line coil can be used as receive-only or transceiver RF coil depending on the targeted application. Our analysis reveals that a decoupling network is only required for the first operating mode. The second operating mode has a high impedance characteristic and can be used to design a tight-fitting transceiver array (Supplementary Figs. 1 and 2). In a multichannel RF array of received-only coils, overlapping decoupling technique combined with low impedance preamplifier [83] can be used to assure sufficient decoupling at both frequencies between adjacent elements (Supplementary Figs. 3 and 4). In other applications where the proposed design is used as building block for transceiver RF array, the magnetic wall decoupling technique, or induced current elimination (ICE) [51] can be used to suppress electromagnetic coupling between adjacent elements (Supplementary Figs. 5 and 6). It's worth stating that different techniques can also be used to mitigate the electromagnetic coupling between neighboring elements.

The sensitivity of the frequency response (S-parameter) of the proposed dual-tuned CTL relative to the diameter of the conductor (inner/outer) and the PTFE (polytetrafluoroethylene) insulation dielectric material is investigated due to uncertainties of the propriety of the commercial coaxial cable. A parametric study using full wave simulation is provided to show the variation of scattering parameter of the proposed dual tuned RF coil relative to the diameter of the conductor (inner/outer) and the PTFE insulation dielectric material of the commercial semi-flexible "RG-405 .086" coaxial cable (Supplementary Fig. 7) used for the coils fabrication. The proposed dual-tuned ($^1$H/$^{13}$C) coaxial-transmission-line RF coil is used as the nominal design for the parametric study. The variation of the inner-conductor diameter $\emptyset$ ($i.c$) from 0.3 to 0.7 mm (while other parameters are fixed) shows a deviation of the dual-band operation including the resonant frequencies as well as the port impedance matching [Supplementary Fig. 8(a)]. It's important to specify that the diameter of the inner conductor is the same as the inner diameter of the PTFE insulation dielectric. Then, the uncertainty of the outer diameter of the PTFE insulation dielectric $\emptyset$ ($i.d$) (from 1.5 to 2mm) is also sensitive to the resonant frequencies and the port input impedance [Supplementary Fig. 8(b)]. Finally, the variation of the outer-conductor diameter $\emptyset$ ($o.c$) from 2 to 3 mm has no impact on the frequency response of the dual-tuned RF coil [Supplementary Fig. 8(c)]. This parametric study reveals that the proposed design is sensitive to the size (cross section) of the PTFE insulation dielectric. The sensitivity is more pronounced at the higher operation frequency; however, this variation can be adjusted by the tuning/matching lumped components.

The characteristics of our proposed dual-tuned ($^1$H/$^{13}$C) coaxial transmission line coil are compared with a conventional single dual-tuned loop coil in term RF transmit/receive ($B_1^+/B_1^-$) field efficiency and quality factor Q. The single loop dual-tuned ($^1$H/$^{13}$C) is based on the "two-pole" technique using an LC trap circuit [64] implemented to design a dual-tuned single-port surface coil. Both designs are loaded with a cylindrical phantom with conductivity value $\sigma$ = 0.6 S/m and the permittivity value $\varepsilon_r = 50$. The diameter and length of the cylindrical phantom are set to 20 cm and 20 cm, respectively and placed 0.5 cm from the RF coils [Supplementary Fig. 9 (a) and (b)]. From the scattering parameters [Supplementary Fig. 9 (c)], it can be seen that low Q is a potential drawback of the coaxial transmission line coil. The simulated loaded quality factors at $^{13}$C and $^1$H of the conventional dual-tuned surface loop coil are 75 and 51 respectively, while the ones evaluated for the proposed coil are 30 and 26, respectively. However, in term of transmit/receive ($B_1^+/B_1^-$) field efficiency both designs exhibit similar $B_1$ field distribution and strength in in the central axial slice of the phantom [Supplementary Fig. 10] at both operating bands.

## V. Conclusion

We have introduced a dual-tuned coaxial transmission line RF coil with independent tuning capability. The analysis based on full-wave simulation and experiment has demonstrated how both resonant frequencies can be controlled individually by simply varying the constituent of the design parameters. A broadband tuning range is obtained, which can cover the proton (high-frequency mode) and most of X-nuclei (low-frequency mode) at 7T. This design characteristic has enabled the implementation of a dual-tuned $^1$H/$^{13}$C for hyperpolarized carbon-13 magnetic resonance imaging ($^{13}$C- HPMRI) and $^1$H/$^2$H for deuterium metabolic imaging ($^2$H-DMI), both useful for studying tissue metabolism. Due to the use of distributed circuit of coaxial transmission line, this new design possesses unique features, including high-frequency operation capability, and doesn't require any decoupling network between the two nuclear channels, which enable highly sensitive detection of MR signals from both proton and heteronuclear signal with similar $B_1$ field distribution. This modest single-coil design lessens the design complexity of dual-tuned RF coils and, more importantly, is a valuable building block of dual-tuned multichannel RF arrays. The successful development of this new dual-tuned RF coil technique would provide a tangible and efficient tool for ultrahigh field metabolic MR imaging.

ACKNOWLEDGMENT

This work is supported in part by the NIH under a BRP grant U01 EB023829 and SUNY Empire Innovation Professorship Award.